\documentclass[aps,
 preprint,
 preprintnumbers,
 preprintnumbers,
 nobibnotes,
 bibnotes,
 amsmath,amssymb,
 aps,
 floatfix,
]{revtex4-1}

\usepackage{graphicx}
\usepackage{dcolumn}
\usepackage{bm}
\usepackage{natbib}
\usepackage{xcolor}

\begin{document}

\title{Coexistence of polar displacements and conduction in doped
ferroelectrics: an \textit{ab initio} comparative study}

\author{Chengliang Xia$^{1,2}$, Yue Chen$^{1, *}$ and Hanghui Chen$^{2,3,}$}
\email{Correspondence to hanghui.chen@nyu.edu or yuechen@hku.hk.}
\affiliation{
  $^1$Department of Mechanical Engineering, The University of Hong Kong, Pokfulam Road, Hong Kong SAR, China \\
  $^2$NYU-ECNU Institute of Physics, NYU Shanghai, Shanghai, 200062, China\\
  $^3$Department of Physics, New York University, New York  10003, USA\\
}


\begin{abstract}  
Polar metals are rare because free carriers in metals screen
electrostatic potential and eliminate internal dipoles. Degenerate
doped ferroelectrics may create an approximate polar metallic phase.
We use first-principles calculations to investigate $n$-doped
LiNbO$_3$-type oxides (LiNbO$_3$ as the prototype) and compare to
widely studied perovskite oxides (BaTiO$_3$ as the prototype). In the
rigid-band approximation, substantial polar displacements in $n$-doped
LiNbO$_3$ persist even at 0.3~$e$/f.u. ($\simeq$ 10$^{21}$ cm$^{-3}$),
while polar displacements in $n$-doped BaTiO$_3$ quickly get
suppressed and completely vanish at 0.1~$e$/f.u. Furthermore, in
$n$-doped LiNbO$_3$, Li-O displacements decay more slowly than Nb-O
displacements, while in $n$-doped BaTiO$_3$, Ba-O and Ti-O
displacements decay approximately at the same rate. Supercell
calculations that use oxygen vacancies as electron donors support the
main results from the rigid-band approximation and provide more detailed
charge distributions. Substantial cation displacements are observed
throughout LiNbO$_{3-\delta}$($\delta = 4.2\%$), while cation
displacements in BaTiO$_{3-\delta}$($\delta = 4.2\%$) are almost
completely suppressed. We find that conduction electrons in
LiNbO$_{3-\delta}$ are not as uniformly distributed as in
BaTiO$_{3-\delta}$, implying that the rigid-band approximation should
be used with caution in simulating electron doped LiNbO$_3$-type
oxides. Our work shows that polar distortions and
conduction can coexist in a wide range of electron concentration in
$n$-doped LiNbO$_3$, which is a practical approach to generating an
approximate polar metallic phase. Combining doped ferroelectrics and
doped semiconductors may create new functions for devices.

\end{abstract}

\pacs{Valid PACS appear here}
\maketitle


\section{\label{sec:level1}Introduction}

Polar metals are materials that are characterized by the absence of
inversion symmetry and the presence of intrinsic conduction due to
partial band
occupation~\cite{puggioni2014designing,cao2018artificial,balachandran2017learning,page2008local,bock2015metallic}. They
are rare in solids because free carriers can screen electrostatic
potential and eliminate internal dipoles that arise from asymmetric
charge
distributions~\cite{mccarty2008electrostatic,rabe2007physics,mooradian1966observation,bowlan2010electric,hill2000there}. Anderson
and Blount predicted in 1965 that polar metals can
exist~\cite{anderson1965symmetry}, and the recent experimental
confirmation of LiOsO$_3$ as the first polar metal has stimulated
intensive theoretical and experimental
research~\cite{shi2013ferroelectric,xiang2014origin,giovannetti2014dual,liu2015metallic,aulesti2018APL,vecchio2016electronic,sim2014first,padmanabhan2018linear,fei2018ferroelectric,kim2016polar}.

However, the above definition of a polar metal (absence of inversion
symmetry and presence of conduction) excludes degenerately doped
insulating ferroelectrics~\cite{kim2016polar}.  Electron doped
perovskite ferroelectric compounds $AB$O$_3$ (BaTiO$_3$ as the
prototype) have been widely studied both in theory and in
experiment~\cite{Jeong2011,hwang2010doping,kolodiazhnyi2010persistence,kolodiazhnyi2008insulator,wang2012ferroelectric,raghavan2016probing,won2011diode}. First-principles
calculations show that cation displacements and conduction can coexist
in $n$-doped BaTiO$_3$ up to a critical concentration of $0.1e$ per
formula~\cite{wang2012ferroelectric}. This indicates that even with
long-range Coulomb interaction screened by free
electrons~\cite{bohm1953collective,cochran1960w,galitski2004universal},
a short-range portion of Coulomb force with an interaction range of
the order of the lattice constant is sufficient to induce
ferroelectric instability in
BaTiO$_3$~\cite{wang2012ferroelectric,zhang2006ferroelectric,cohen1992origin}. Experimentally,
there are contradictory results: Ref.~\cite{hwang2010doping, kolodiazhnyi2010persistence} show that in oxygen-reduced
BaTiO$_{3-\delta}$, polar displacements can co-exist with conduction
and do not vanish until a critical concentration of $1.9\times10^{21}$
cm$^{-3}$, which is consistent with first-principles
calculations~\cite{wang2012ferroelectric}.  However, a neutron
diffraction study on $n$-doped BaTiO$_3$ found phase separation in
which ferroelectric displacements only exist in an insulating region,
which is spatially separated by nonpolar metallic
regions~\cite{Jeong2011}.  On the other hand, while electron-doped
LiNbO$_3$-type ferroelectric oxides (LiNbO$_3$ as the prototype) have
been investigated in the literature, the focus has been on electronic
structure and the optical
property~\cite{furukawa2001green,nataf2016low,nakamura2002periodic,noguchi2016electronic}. The
structural property and the possible co-existence of polar displacements
with conduction have received little attention.

In this work, we use first-principles calculations to do a comparative
study on doping effects in insulating ferroelectrics. We compare the
aforementioned two important classes of ferroelectrics: one is
perovskite oxides (BaTiO$_3$ as the prototype) and the other is
LiNbO$_3$-type oxides (LiNbO$_3$ as the prototype). We find different
behaviors in these two materials upon electron doping. In the rigid-band
approximation, cation displacements in $n$-doped BaTiO$_3$ quickly get
suppressed and completely disappear at a critical doping of
0.1~$e$/f.u., while substantial cation displacements are found in
$n$-doped LiNbO$_3$ up to an electron concentration of
0.3~$e$/f.u.. Moreover, Li-O displacements decay more slowly than Nb-O
displacements. With an electron doping of 0.3~$e$/f.u.,  Nb-O
displacements are reduced by about 50\% from the undoped value, while
Li-O displacements are reduced by only about 10\%.  This is different
from $n$-doped BaTiO$_3$ in which both Ba-O and Ti-O displacements
decay approximately at the same rate. This indicates that Li-O
displacements are more persistent than Nb-O displacements in
a metallic environment.
Supercell calculations that use oxygen vacancy as electron donors
support the main results from the rigid-band approximation and provide
more detailed charge distributions. We find that in $n$-doped
LiNbO$_3$, conduction electrons are not as uniformly distributed as in
$n$-doped BaTiO$_3$, but
substantial cation displacements are found throughout $n$-doped LiNbO$_3$. Using supercell
calculations, we also compute the formation energy of oxygen vacancies. The
formation energy of oxygen vacancies in $n$-doped LiNbO$_3$ is about 1
eV higher than that in $n$-doped BaTiO$_3$, which is reasonable considering
the fact that the band gap of LiNbO$_3$ is about 1 eV larger than that
of BaTiO$_3$.

The paper is organized as follows. In Section II we provide computation
details. We present the main results (rigid-band calculations and
supercell calculations) in Section III. We conclude in Section IV. 

\section{\label{sce.level1}Computational details}

We perform density functional (DFT) calculations
~\cite{hohenberg1964inhomogeneous,kohn1965self}, as implemented in the
Vienna Ab-initio Simulation Package
(VASP)~\cite{payne1992iterative,kresse1996efficient}.  We employ a local
density approximation (LDA)
pseudopotential~\cite{ceperley1980ground}. We also test our key
results using a revised Perdew-Burke-Ernzerhof generalized gradient
approximation (PBEsol)~\cite{perdew2008jp}.
The key results do
not qualitatively change with different exchange correlation
functionals.  We set an energy cutoff of 600 eV. Charge
self-consistent calculations are converged to 10$^{-5}$ eV. Both cell
and internal coordinates are fully relaxed until each force component
is smaller than 10 meV/\AA\ and the stress tensor is smaller than 1 kbar.

For pristine bulk calculations, we use a tetragonal cell (5-atom) to study
BaTiO$_3$ and find that $a$ = 3.95 \AA~and $c/a$ = 1.01; we
use a hexagonal cell (30-atom) to study $R3c$ LiNbO$_3$ and find that $a$ =
5.09 \AA~and $c$ = 13.81 \AA. Both of them are in good agreement with
previous studies~\cite{zhang2017comparative}.

To simulate doping effects, we use both the rigid-band approximation and
supercell calculations. In rigid-band modeling, we study a 5-atom cell
BaTiO$_3$ (tetragonal $P4mm$ and cubic $Pm\bar{3}m$) and a 30-atom cell
LiNbO$_3$ (non-centrosymmetric $R3c$ and centrosymmetric $R\bar{3}c$).
We use a Monkhorst-Pack $k$-point sampling of $14\times14\times14$ for
BaTiO$_3$ and $8\times8\times3$ for LiNbO$_3$. In supercell
calculations, we use a 119-atom cell for both BaTiO$_3$ and LiNbO$_3$
(oxygen vacancy concentration of 4.2\%/f.u. and nominally electron
doping of 0.084~$e$/f.u.). The supercells for BaTiO$_3$ and LiNbO$_3$
are shown in Fig.~\ref{fig:BaTiO3-supercell} and
Fig.~\ref{fig:LiNbO3-supercell}. We use a Monkhorst-Pack $k$-point
sampling of $8\times8\times8$ in supercell calculations.

In our supercell calculations, we remove one (charge neutral) oxygen
atom in LiNbO$_3$ supercells of different sizes to simulate different
oxygen vacancy concentrations. The supercell with oxygen vacancies is
charge-neutral, and we fully relax the structure (both lattice
constants and internal coordinates) to get the ground state property.

We check a higher energy cutoff (750 eV) and a denser $k$-point
sampling, and we do not find any significant changes in the key
results.

\section{Results and discussion}

\subsection{Rigid-band calculations}

\begin{figure}[t!]
\includegraphics[scale=0.5]{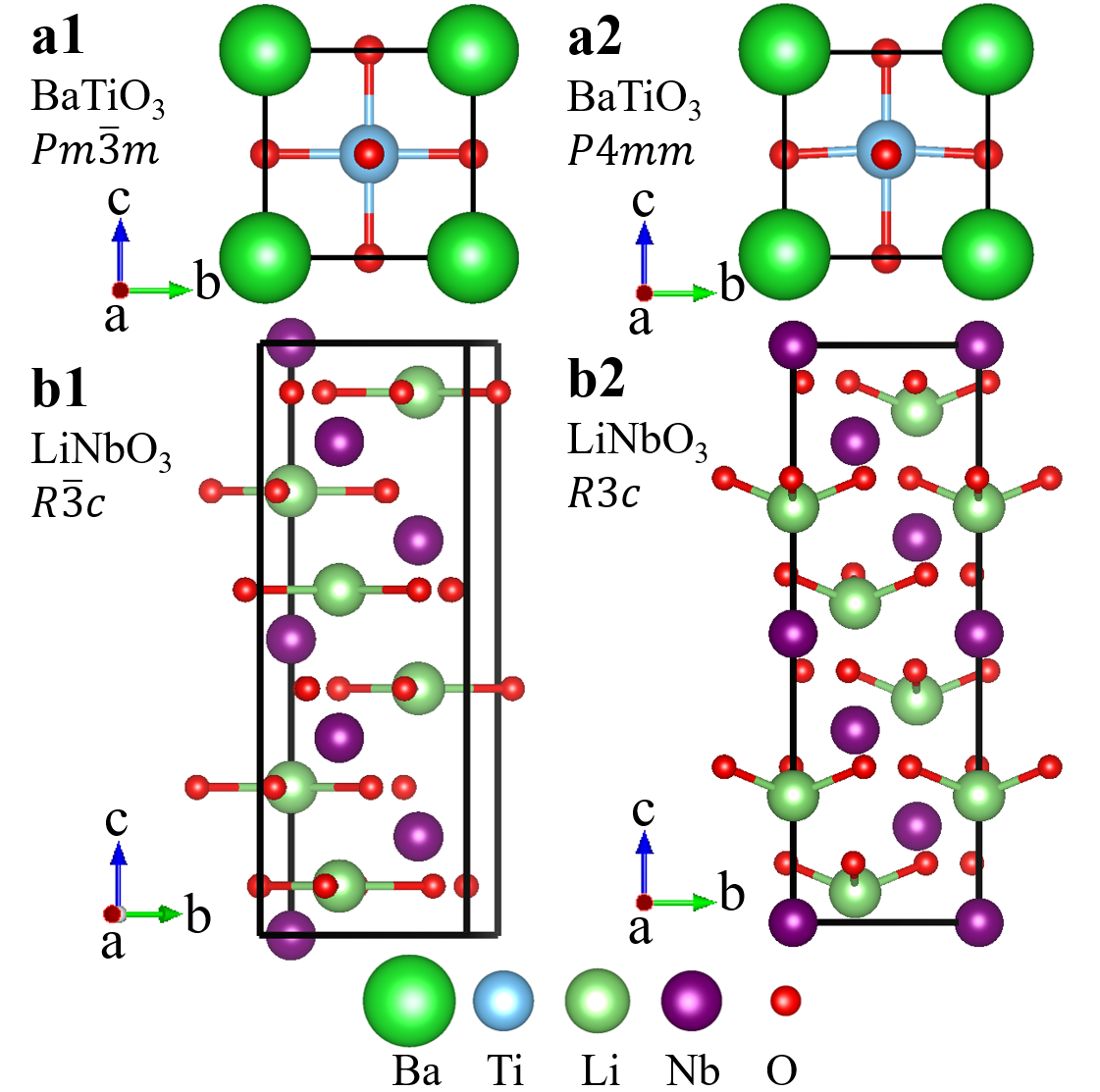}
\caption{Atomic structures of BaTiO$_3$ and
  LiNbO$_3$. Panels \textbf{a1)} and \textbf{a2)} show cubic
  $Pm\overline{3}m$ and tetragonal $P4mm$ structures of BaTiO$_3$,
  respectively. Panels \textbf{b1)}
  and \textbf{b2)} show centrosymmetric $R\overline{3}c$ and non-centrosymmetric $R3c$
  structures of LiNbO$_3$, respectively.}
\label{fig:crystal}
\end{figure}

In the rigid-band approximation, materials are pristine and extra
electrons are added to the system with the same amount of uniform
positive charges in the background.  Fig.~\ref{fig:crystal} shows the
crystal structures of pristine BaTiO$_3$ and LiNbO$_3$, which are used
in rigid-band modeling. Panels \textbf{a1} and \textbf{a2} show the
crystal structure of cubic BaTiO$_3$ (space group $Pm\overline{3}m$)
and tetragonal BaTiO$_3$ (space group
$P4mm$). Insulating ferroelectrics have a spontaneous
polarization~\cite{abrahams1968atomic}. However, in doped
ferroelectrics, partially filled bands may lead to conduction and
polarization becomes
ill-defined~\cite{wang2010zero,stengel2011band,fujioka2015ferroelectric}.
Therefore, we use cation displacements to characterize the extent of
being ``polar''.  In $n$-doped BaTiO$_3$, we calculate both Ba-O and
Ti-O cation displacements along the $c$-axis, denoted by
$\delta_{\rm{Ba-O}}$ and $\delta_{\rm{Ti-O}}$, as a function of
electron concentration. $\delta_{\textrm{Ba}-\textrm{O}}$ and
$\delta_{\textrm{Ti}-\textrm{O}}$ are explicitly shown in Fig. S1
in the Supplementary Materials.
Panels \textbf{b1} and \textbf{b2} show the crystal
structure of centrosymmetric LiNbO$_3$ (space group $R\overline{3}c$)
and non-centrosymmetric LiNbO$_3$ (space
group $R3c$). We calculate both Li-O and Nb-O displacements
$\delta_{\textrm{Li}-\textrm{O}}$ and
$\delta_{\textrm{Nb}-\textrm{O}}$. In the centrosymmetric structure
$R\overline{3}c$, each Li atom is surrounded by three oxygen atoms and
all these four atoms form a plane that is perpendicular to
$c$-axis. In the non-centrosymmetric structure $R3c$, the three oxygen
atoms still form a plane that is perpendicular to the $c$-axis but Li
atom deviates from that plane. The distance between Li atom and the
plane that the three oxygen atoms form is defined as the Li-O
displacement $\delta_{\textrm{Li}-\textrm{O}}$. In the centrosymmetric
structure $R\overline{3}c$, each Nb atom is at the center of an
NbO$_6$ oxygen octahedron. In the non-centrosymmetric
structure $R3c$, Nb atoms move off the center of the NbO$_6$ oxygen
octahedron. The distance between the Nb position and the center
of the oxygen octahedron in the $R3c$ structure is defined as
$\delta_{\textrm{Nb}-\textrm{O}}$. $\delta_{\textrm{Li}-\textrm{O}}$
and $\delta_{\textrm{Nb}-\textrm{O}}$ are explicitly shown in Fig. S2
in the Supplementary Materials.

Fig.~\ref{fig:ferro-rigid} summarizes the key results from rigid-band
calculations.  Panel \textbf{a} shows the cation displacements of
tetragonal BaTiO$_3$ (space group $P4mm$) and non-centrosymmetric
LiNbO$_3$ (space group $R3c$). In the undoped case,
$\delta_{\textrm{Ba}-\textrm{O}} = 0.077$~\AA~and
$\delta_{\textrm{Ti}-\textrm{O}} = 0.099$~\AA~in BaTiO$_3$, and
$\delta_{\textrm{Li}-\textrm{O}} = 0.723$~\AA~and
$\delta_{\textrm{Nb}-\textrm{O}} = 0.261$~\AA~in LiNbO$_3$, both of
which are in good agreement with previous calculations and
experiments~\cite{hellwege1981landolt,prokhorov1990physics,nahm2001microscopic,rabe2007physics}.
Upon doping, all of the cation displacements decrease with increasing electron
concentration. $\delta_{\textrm{Ba}-\textrm{O}}$ and
$\delta_{\textrm{Ti}-\textrm{O}}$ in $n$-doped BaTiO$_3$ vanish at
$n_c \simeq 0.1~e$/f.u., which is consistent with previous
calculations~\cite{wang2012ferroelectric}. However,
$\delta_{\textrm{Li}-\textrm{O}}$ and
$\delta_{\textrm{Nb}-\textrm{O}}$ in $n$-doped LiNbO$_3$ persist up to
0.3~$e$/f.u.. Furthermore, in $n$-doped BaTiO$_3$,
$\delta_{\textrm{Ba-O}}$ and $\delta_{\textrm{Ti-O}}$ decay at
approximately the same rate. But in $n$-doped LiNbO$_3$,
$\delta_{\textrm{Li-O}}$ decays more slowly than
$\delta_{\textrm{Nb-O}}$.
With 0.3~$e$/f.u. electron doping, Nb-O
displacements are reduced by about 50\% from the undoped value, while
Li-O displacements are reduced by only about 10\%. This indicates that
the off-center movements of Li atoms are very robust and
more persistent than Nb-O displacements in a metallic environment.
This helps to create an approximate polar metallic phase when
LiNbO$_3$ is electron doped.

Panel \textbf{b} of Fig.~\ref{fig:ferro-rigid} shows the energy
difference between the centrosymmetric structure and the
non-centrosymmetric structure of BaTiO$_3$ and
LiNbO$_3$. Specifically, for BaTiO$_3$ $\Delta E = E(Pm\overline{3}m)
- E(P4mm)$ and for LiNbO$_3$ $\Delta E = E(R\overline{3}c) -
E(R3c)$. $\Delta E > 0$ indicates that the non-centrosymmetric
structure is favored. In the undoped case, the non-centrosymmetric
structure is favored in both BaTiO$_3$ and LiNbO$_3$, i.e. they are
both ferroelectric. Upon doping, BaTiO$_3$ is polar till $n_c \simeq
0.1~e$/f.u., consistent with the critical concentration found for
$\delta_{\textrm{Ba}-\textrm{O}}$ and
$\delta_{\textrm{Ti}-\textrm{O}}$. For $n$-doped LiNbO$_3$, $\Delta E$
quickly decreases but it stays positive (up to 0.3~$e$/f.u.).  This is
consistent with $\delta_{\textrm{Li-O}}$ and $\delta_{\textrm{Nb-O}}$,
which do not vanish with electron doping (up to 0.3~$e$/f.u.).

Panel \textbf{c} of Fig.~\ref{fig:ferro-rigid} shows the zone-center
phonon frequency of the ferroelectric mode for centrosymmetric
BaTiO$_3$ (space group $Pm\overline{3}m$) and LiNbO$_3$
(space group $R\overline{3}c$). For cubic $Pm\overline{3}m$ BaTiO$_3$, the
ferroelectric mode has imaginary phonon frequency with
small electron doping, indicating ferroelectric instability. Around
the critical doping of $n_c \simeq 0.1~e$/f.u. the phonon frequency of
the ferroelectric mode becomes positive and the cubic structure is
stabilized.  For centrosymmetric $R\overline{3}c$ LiNbO$_3$, the ferroelectric
mode always has imaginary phonon frequency (up to
0.3~$e$/f.u.), indicating that ferroelectric instability persists in
$n$-doped LiNbO$_3$. For both materials, the phonon property of the
centrosymmetric structures is consistent with the results of the
non-centrosymmetric structures shown in panels \textbf{a} and
\textbf{b}. We note that the magnitude of the imaginary phonon mode indicates how
unstable the high-symmetry structure is subject to a collective
atomic distortion. However, the energy difference
between the distorted and undistorted crystal structures
reflects not only the instability of the high-symmetry structure,
but also other factors. From our calculations, we find that for
BaTiO$_3$, the volume of its undistorted structure is 0.5\% smaller than
that of the distorted structure; in contrast, for LiNbO$_3$,
the volume of its undistorted structure is 1.5\% larger than
that of the distorted structure. The elastic energy change
from the high-symmetry structure to the low-symmetry structure is very
different between BaTiO$_3$ and LiNbO$_3$. This information is embodied
in the total energy difference but is not directly reflected in the
imaginary phonon modes.

We also use the PBEsol functional to check the key results in
Fig.~\ref{fig:ferro-rigid}, and we do not find significant changes (see
Fig. S4 in the Supplementary Materials). Our finding that polar
displacements of LiNbO$_3$ are persistent in the presence of
conduction electrons is related to the fact that LiNbO$_3$ is
hyperferroelectric~\cite{garrity2014hyperferroelectrics,li2016origin}, i.e.
a ferroelectric material whose polarization does not get suppressed by
depolarization fields. This implies that doping a
hyperferroelectric material is a viable approach to generating an
approximate polar metallic phase.

\begin{figure}[t]
\includegraphics[scale=0.38]{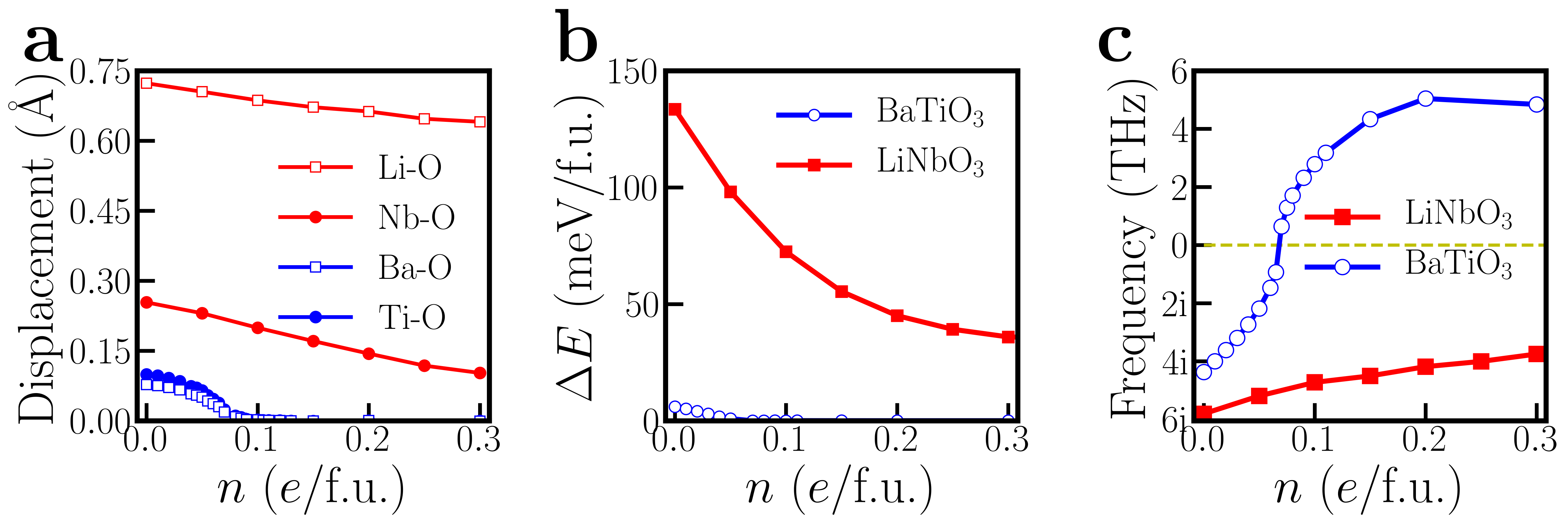}
\caption{Panel \textbf{a)} Ba-O, Ti-O, Li-O and Nb-O displacements in
  non-centrosymmetric structures of BaTiO$_3$ and LiNbO$_3$ as a
  function of electron doping.  Panel \textbf{b)} energy difference
  between the centrosymmetric and the non-centrosymmetric structures
  of BaTiO$_3$ and LiNbO$_3$ as a function of electron doping.  Panel
  \textbf{c)} phonon frequency of the zone-center ferroelectric mode
  of cubic BaTiO$_3$ and $R\overline{3}c$ LiNbO$_3$ as a function of
  electron doping.}
\label{fig:ferro-rigid}
\end{figure}

\begin{figure}[t]
\includegraphics[scale=0.25]{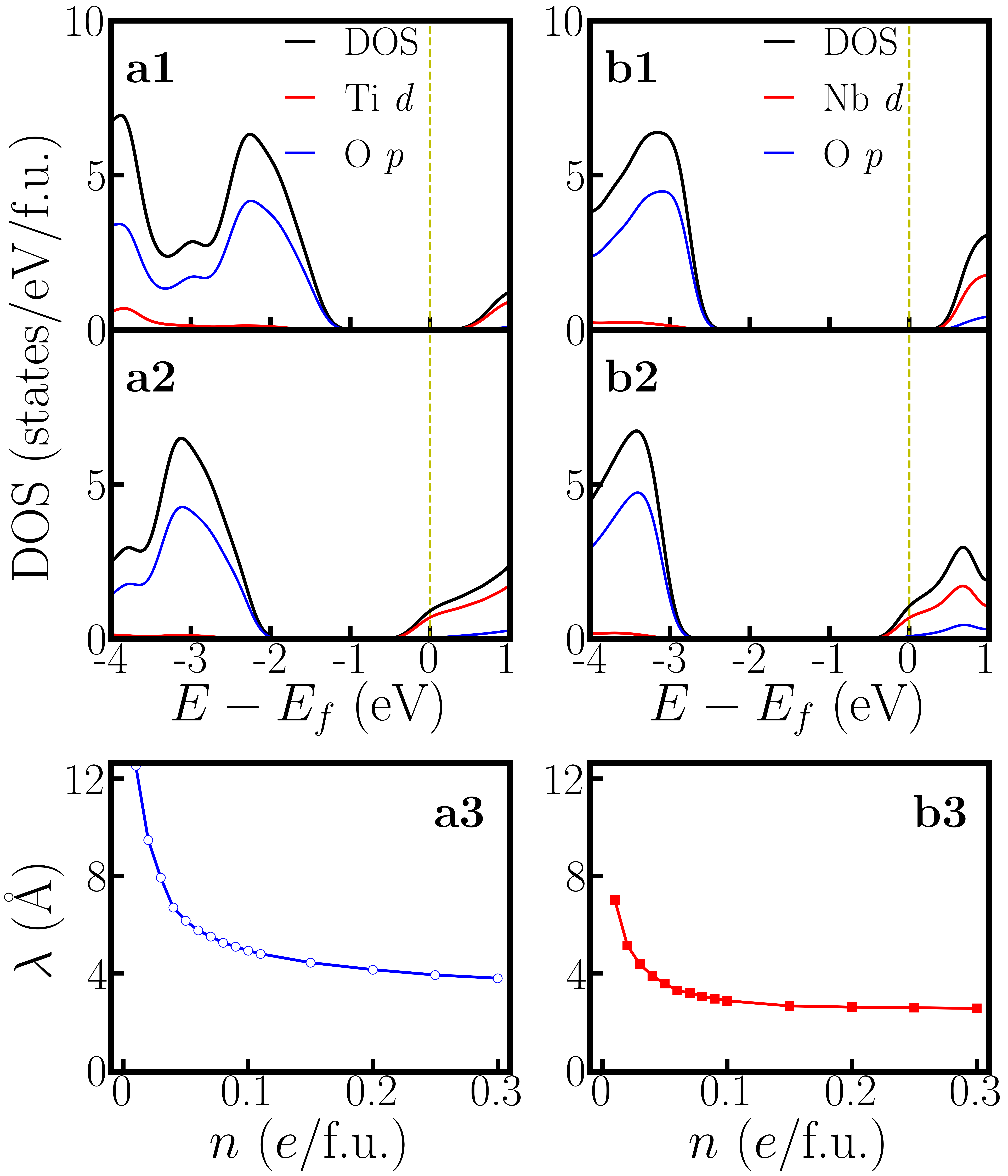}
\caption{Panel \textbf{a1)} density of states of undoped BaTiO$_3$.
  Panel \textbf{a2)} density of states of doped BaTiO$_3$ with 0.2~
  $e$/f.u. doping. The black, red and blue are total, Ti-$d$ and O-$p$ projected
  densities of states, respectively. 
  Panel \textbf{b1)} density of states of undoped
  LiNbO$_3$. Panel \textbf{b2)} density of states of doped LiNbO$_3$
  with 0.2~$e$/f.u. doping. The black, red and blue are total, Nb-$d$ and O-$p$
  projected densities of states respectively.
  Panel \textbf{a3)} Thomas-Fermi screening length
  $\lambda$ of doped BaTiO$_3$ as a function of electron doping 
  $n$. Panel \textbf{b3)} Thomas-Fermi screening
  length $\lambda$ of doped LiNbO$_3$ as a function of electron doping $n$.}
  \label{fig:ferro-dos}
\end{figure}

Next, we study the electronic structure and screening length obtained from
rigid-band calculations. Panels \textbf{a1} and \textbf{a2} of
Fig.~\ref{fig:ferro-dos} show the density of states of undoped and
doped BaTiO$_3$ (with 0.2~$e$/f.u. doping). With electron doping, the
Fermi level is shifted from the band gap into Ti-$d$ states. Panels
\textbf{b1} and \textbf{b2} of Fig.~\ref{fig:ferro-dos} show the
density of states of undoped and doped LiNbO$_3$ (with
0.2~$e$/f.u. doping). Similarly, with electron doping, the Fermi level
is shifted from the band gap into Nb-$d$ states. With the density of states of
$n$-doped BaTiO$_3$ and $n$-doped LiNbO$_3$, we can estimate the
screening length $\lambda$ by using the Thomas-Fermi
model~\cite{wang2012ferroelectric}:

\begin{equation}
\lambda=\sqrt{\frac{\epsilon}{e^2\times D(E_f)}}
\end{equation}   
where $\epsilon$ is the dielectric constant of undoped materials and
$D(E_f)$ is density of states at Fermi level. For dielectric
constants, we use experimental values $\epsilon \approx 44\epsilon_0$
for BaTiO$_3$~\cite{rupprecht1964dielectric} and $\epsilon \approx
24\epsilon_0$ for LiNbO$_3$~\cite{mansingh1985ac}. Panels \textbf{c}
of Fig.~\ref{fig:ferro-dos} show the screening length of $n$-doped
BaTiO$_3$ and $n$-doped LiNbO$_3$. We find that for both materials
upon electron doping, the screening length is on the order of a
few~\AA. Given an electron concentration, $n$-doped LiNbO$_3$ even has
a screening length slightly smaller than $n$-doped BaTiO$_3$, implying a
stronger screening property. The stronger screening property of
electron doped LiNbO$_3$ is due to the fact that undoped LiNbO$_3$ has
a smaller dielectric constant than that of BaTiO$_3$, while the
density of states at the Fermi level plays a minor role (the ratio of
$D(E_f)_{\textrm{LiNbO}_3}$ to $D(E_f)_{\textrm{BaTiO}_3}$ ranges from
0.9 to 1.1 as $n$ changes from 0 to 0.3~$e$/f.u.).

\subsection{Supercell calculations}

\begin{figure}[t!]
  \includegraphics[height=9.3cm]{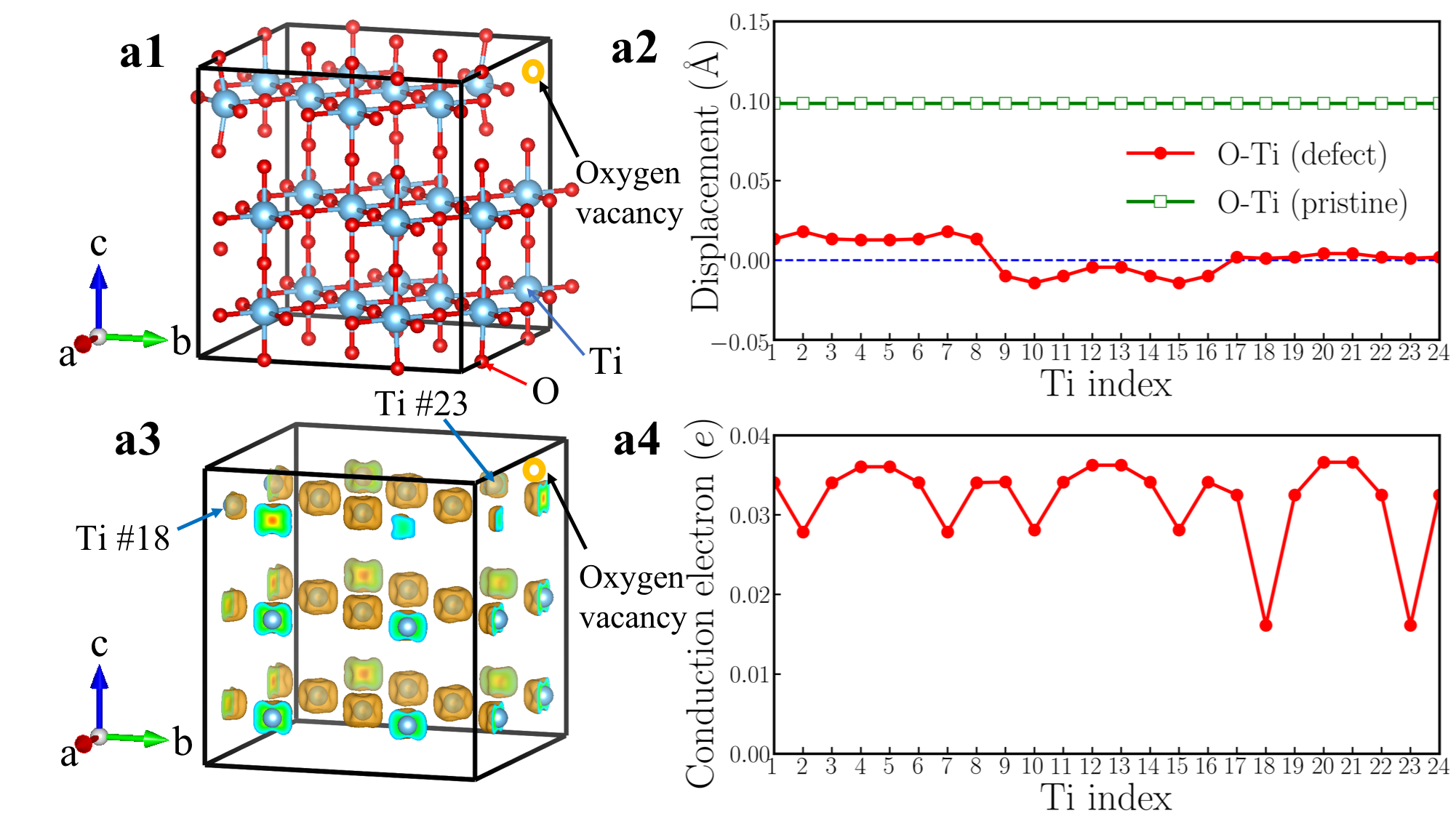}
 \caption{Panel \textbf{a1)} atomic structure of 119-atom
   BaTiO$_{3-\delta}$ supercell that contains an oxygen vacancy. For
   clarity, Ba atoms are not shown. The blue and red balls represent
   Ti and O atoms. Panel \textbf{a2)} polar displacements of each Ti
   atom. The red circles are calculated from the 119-atom cell.  The
   green squares are bulk $\delta_{\textrm{Ti-O}}$ of pristine
   BaTiO$_3$.  Panel \textbf{a3)} an iso-value surface of conduction
   electron distribution in oxygen-reduced BaTiO$_{3-\delta}$ with
   $\delta = 4.2\%$. The two nearest neighbor Ti atoms are highlighted.
   Panel \textbf{a4)} conduction electrons on
   each Ti atom in the 119-atom cell (the number of conduction
   electron on each Ti atom is obtained by integrating Ti-$d$ states
   from band gap to the Fermi level).}
  \label{fig:BaTiO3-supercell}
\end{figure}

Our rigid-band calculations show that polar displacements and
conduction can coexist in both $n$-doped BaTiO$_3$ and $n$-doped
LiNbO$_3$, but the overall polar property (magnitude of polar
displacements, polar instability, etc.) is much more enhanced in
$n$-doped LiNbO$_3$ than in $n$-doped BaTiO$_3$.  However, rigid-band
calculations do not specify the origin of electron doping and also
they imply that all carriers are uniformly distributed. In real
experiments, oxygen vacancies are commonly seen in complex oxides, and
each oxygen vacancy nominally donates two electrons. However, an isolated oxygen
vacancy may form a defect state, which can localize conduction
electrons~\cite{magyari2012first,lin2015localized}.  Some experiments
shows that in oxygen-reduced BaTiO$_{3-\delta}$, phase separation
occurs.  Cation displacements $\delta_{\textrm{Ti}-\textrm{O}}$ only
occur in the insulating region and vanish in the metallic region. The
overall sample may be considered as a mixture of two different
phases~\cite{Jeong2011}.  To test whether the results from the rigid-band calculations remain valid in real materials, we perform supercell
calculations and consider charge neutral oxygen vacancies as the
electron doping source.
 We use a 119-atom cell of BaTiO$_{3-\delta}$
and LiNbO$_{3-\delta}$. In both cases, the oxygen vacancy
concentration is 4.2\%/f.u.. A charge neutral oxygen vacancy donates
two electrons to the system, therefore it is an electron doping of
0.084~$e$/f.u. ($\simeq 1.5\times 10^{21}$ cm$^{-3}$), close to the
critical doping in $n$-doped BaTiO$_3$ obtained from rigid-band
calculations.

Fig.~\ref{fig:BaTiO3-supercell}\textbf{a1} shows the crystal structure
of a 119-atom BaTiO$_{3-\delta}$ supercell that contains one oxygen
vacancy. For clarity, Ba atoms are not explicitly shown and the oxygen
vacancy is highlighted by the orange open circle.
Fig.~\ref{fig:BaTiO3-supercell}\textbf{a2} shows the cation
displacement $\delta_{\textrm{Ti}-\textrm{O}}$ for each Ti atom in the
BaTiO$_{3-\delta}$ supercell (the definition of
$\delta_{\textrm{Ti}-\textrm{O}}$ is identical to that in the rigid-band calculations). Displacements along $c$-axis of each Ba atom are
explicitly shown in Fig. S3 in the Supplementary Materials. We find
that while there is some small variation in
$\delta_{\textrm{Ti}-\textrm{O}}$ due to an inhomogeneous chemical
environment, $\delta_{\textrm{Ti}-\textrm{O}}$ on average is reduced
to zero. For comparison, we also show the
$\delta_{\textrm{Ti}-\textrm{O}}$ in pristine BaTiO$_3$ in panel
\textbf{a2} and the suppression of polar displacements by electron
doping is evident.  Fig.~\ref{fig:BaTiO3-supercell}\textbf{a3} shows
an iso-value surface of conduction electron density in oxygen-reduced
BaTiO$_{3-\delta}$ with $\delta = 4.2\%$. Conduction electrons reside
on Ti atoms.  Because the polar displacements are suppressed and the
material is close to a cubic structure, conduction electrons occupy
three Ti $t_{2g}$ orbitals with approximately equal occupancy. This
leads to an iso-value surface of a dice-like
shape. Fig.~\ref{fig:BaTiO3-supercell}\textbf{a4} shows the number of
conduction electrons on each Ti atom by integrating the Ti-$d$ states
from the band gap to the Fermi level. We find that in the presence of
oxygen vacancy, while there is non-negligible variation in conduction
electron distribution, insulating-metallic phase separation does not
occur in our first-principles calculations. Each Ti atom in the
supercell has a sizable amount of conduction electron. The results of
oxygen-reduced BaTiO$_{3-\delta}$ from supercell calculations are very
consistent with rigid-band calculations. We also use the LDA+$U$ method
and change the supercell size to test the robustness of this conclusion
(see the Supplementary Materials for details). We find that in oxygen-reduced BaTiO$_{3-\delta}$, conduction electrons on each Ti atom are
almost uniformly distributed.

\begin{figure}[t!]
\includegraphics[height=9.3cm]{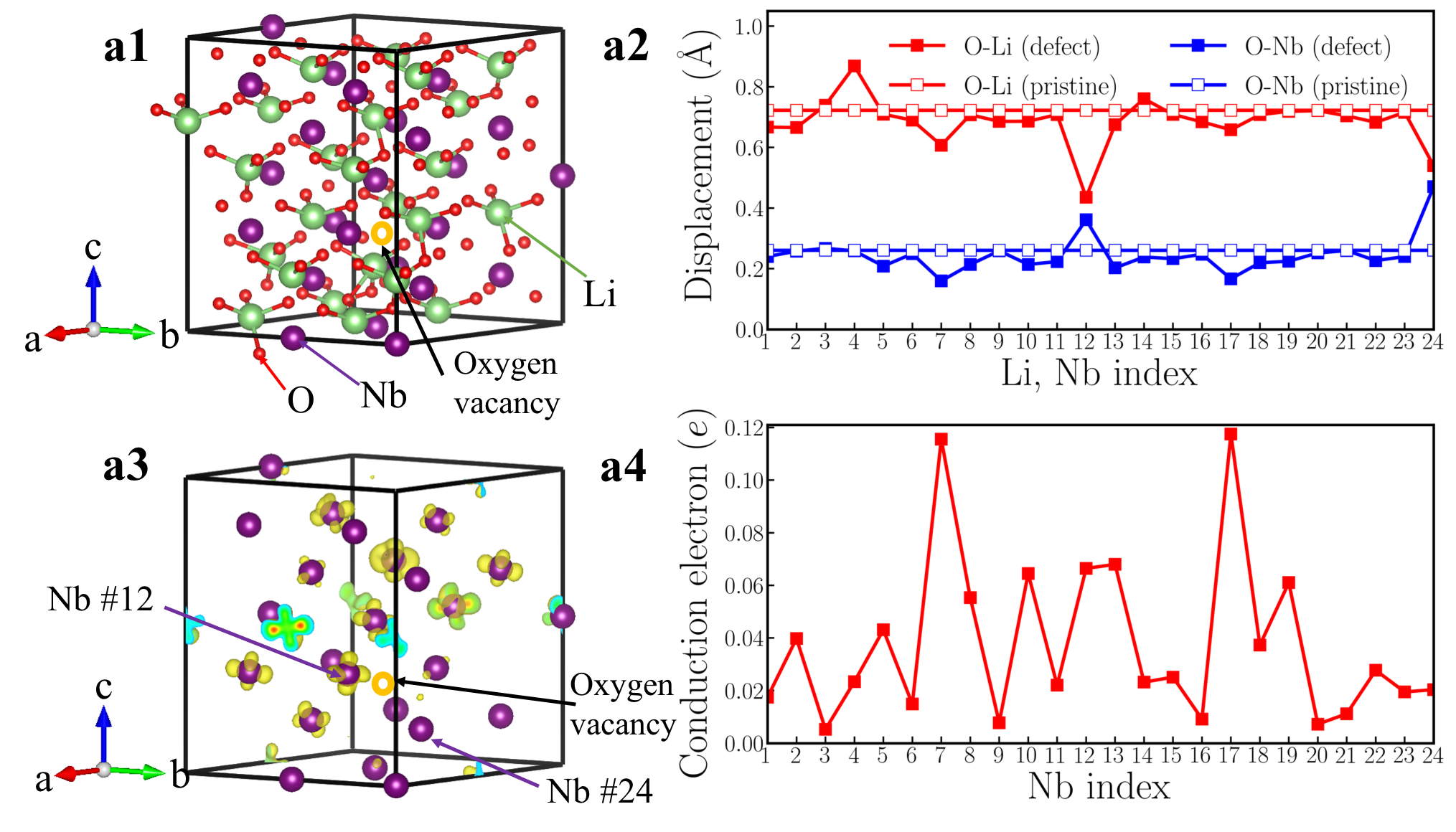}
\caption{Panel \textbf{a1)} atomic structure of 119-atom LiNbO$_3$
  supercell that contains an oxygen vacancy. The green, purple and red
  balls represent Li, Nb and O atoms respectively. The orange circle
  highlights an oxygen vacancy. Panel \textbf{a2)} polar displacements of
  each Li (red square) and Nb atom (blue square). The solid squares
  are calculated from the 119-atom cell.  The open squares are bulk
  $\delta_{\textrm{Li-O}}$ and $\delta_{\textrm{Nb-O}}$ of pristine
  LiNbO$_3$. Panel \textbf{a3)} an iso-value surface of conduction
  electron distribution in oxygen-reduced LiNbO$_{3-\delta}$ with
  $\delta = 4.2\%$. The two nearest neighbor Nb atoms are highlighted.
  Panel \textbf{a4)} conduction electrons on each Nb
  atom in the 119-atom cell (the number of conduction electron on each
  Nb atom is obtained by integrating Nb-$d$ states from band gap to
  the Fermi level).}

\label{fig:LiNbO3-supercell}
\end{figure}

However, supercell calculations of oxygen-reduced LiNbO$_{3-\delta}$
show more complicated results than rigid-band calculations.
Fig.~\ref{fig:LiNbO3-supercell}\textbf{a1} shows the crystal structure
of a 119-atom LiNbO$_3$ supercell that contains one oxygen vacancy. The
oxygen vacancy is highlighted by the orange open circle.
Fig.~\ref{fig:LiNbO3-supercell}\textbf{a2} shows
$\delta_{\textrm{Li}-\textrm{O}}$ for each Li atom and
$\delta_{\textrm{Nb}-\textrm{O}}$ for each Nb atom in the
LiNbO$_{3-\delta}$ supercell (the definitions of
$\delta_{\textrm{Li}-\textrm{O}}$ and
$\delta_{\textrm{Nb}-\textrm{O}}$ are identical to those in the rigid-band calculations). We find that while there is
non-negligible variation in $\delta_{\textrm{Li}-\textrm{O}}$ and
$\delta_{\textrm{Nb}-\textrm{O}}$, the cation displacements on each Li
and Nb atoms are non-zero throughout the supercell. For comparison, we
also show the $\delta_{\textrm{Li}-\textrm{O}}$ and
$\delta_{\textrm{Nb}-\textrm{O}}$ in pristine LiNbO$_3$ in panel
\textbf{a2}. We find that in the presence of oxygen vacancy in
LiNbO$_{3-\delta}$, the cation displacements are still substantial,
compared to undoped LiNbO$_3$.

The average value of $\delta_{\textrm{Li}-\textrm{O}}$ is 0.68~\AA~and
the average value of $\delta_{\textrm{Nb}-\textrm{O}}$ is 0.24~\AA,
both of which are close to the results of rigid-band calculations with
the same electron concentration (in the rigid-band approximation,
$\delta_{\textrm{Li}-\textrm{O}}$ is 0.69~\AA~ and
$\delta_{\textrm{Nb}-\textrm{O}}$ is 0.21~\AA).
Fig.~\ref{fig:LiNbO3-supercell}\textbf{a3} shows an iso-value surface
of conduction electron distribution in oxygen-reduced
LiNbO$_{3-\delta}$ with $\delta = 4.2\%$. Different from oxygen-reduced BaTiO$_{3-\delta}$ which has almost uniform conduction
electron distribution, some of Nb atoms have negligible conduction
electrons, indicating that those Nb sites are almost insulating. This
phenomenon can be more clearly seen from
Fig.~\ref{fig:LiNbO3-supercell}\textbf{a4}, which shows the number of
conduction electrons on each Nb site by integrating Nb-$d$ states from
the band gap to the Fermi level. Nb \#3, \#9 and \#20 have less than
0.01 $e$ per atom, while Nb \#7 and \#17 have more than 0.12 $e$ per
atom. With electron doping in LiNbO$_3$, the conduction electrons on Nb sites are far from uniformly distributed. 
This implies that the charge disproportionation of conduction electrons on Nb atoms occurs in real materials. Such charge disproportion can be suppressed with a higher electron concentration. We
calculated oxygen-reduced LiNbO$_{3-\delta}$ with $\delta = 8.4\%$.
Cation displacements and conduction electron distribution on Nb-$d$
states are shown in Fig. S9 in the Supplementary Materials. While
there is non-negligible variation in the electron distribution, all
Nb atoms have conduction electrons with a higher concentration of
oxygen vacancies, as expected. 

\begin{figure}[t!]
\includegraphics[scale=0.7]{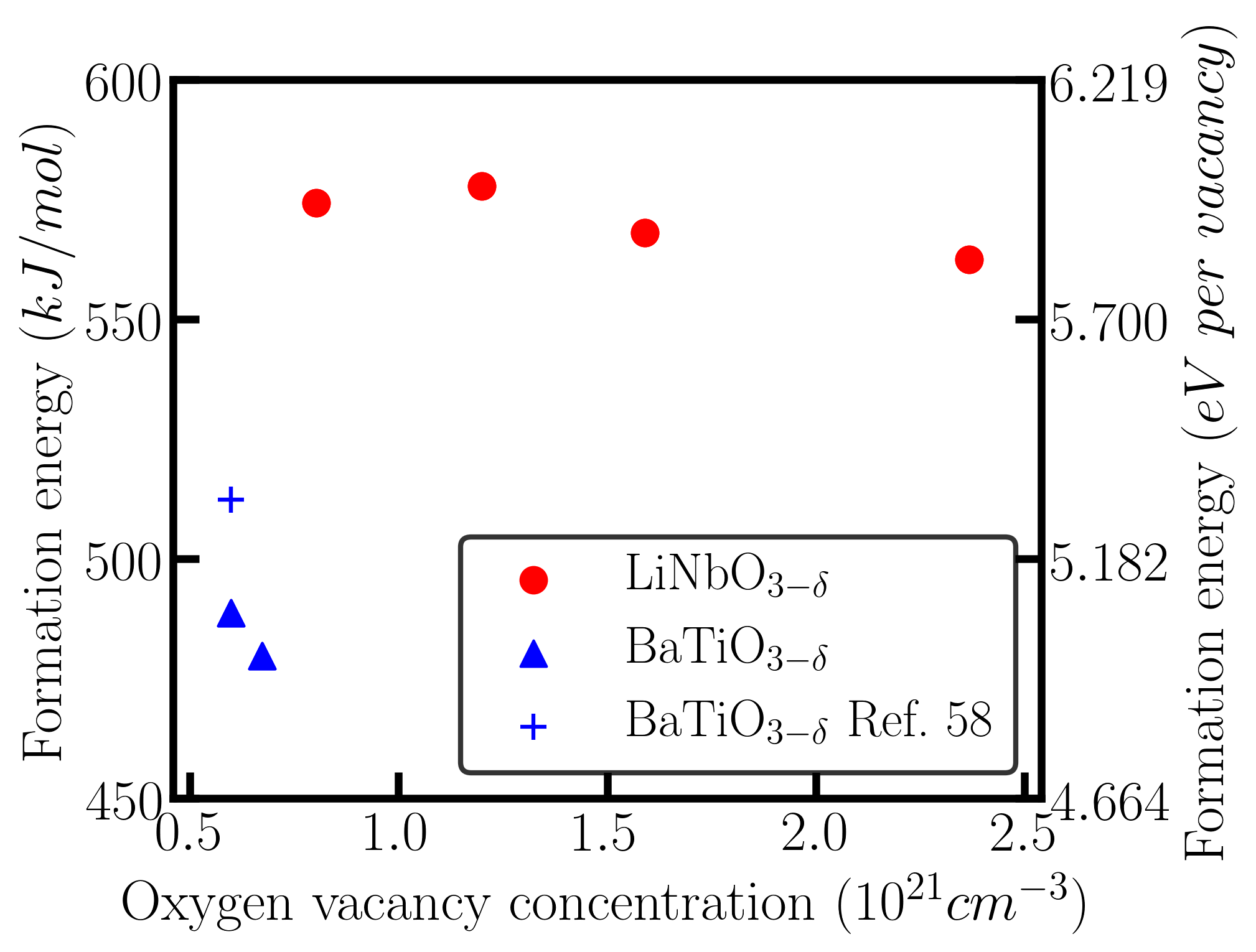}
\caption{Oxygen vacancy formation energy in oxygen-reduced
  BaTiO$_{3-\delta}$ (blue triangles) and in oxygen-reduced
  LiNbO$_{3-\delta}$ (red circles) as a function of oxygen vacancy
  concentration. The blue cross is a reference value from previous
  works~\cite{choi2011electronic}.}
\label{fig:fm}
\end{figure}

Finally we estimate the formation energy of a charge-neutral oxygen
vacancy in BaTiO$_{3-\delta}$ and in LiNbO$_{3-\delta}$ using a
supercell calculation. We study a charge neutral
  oxygen vacancy by removing an oxygen atom from a charge neutral
  supercell. The definition of formation energy of a charge-neutral
oxygen vacancy in the oxygen rich limit is:

\begin{equation}
\Delta E_{\textrm{O}}^f= E_T(V_{\textrm{O}})-E_{T0}+\frac{1}{2}E_{\textrm{O}_2}
\end{equation} 
where $E_T(V_{\textrm{O}})$ is the total energy of a supercell with one oxygen
vacancy, and $E_{T0}$ is the total energy of a pristine supercell.
$E_{\textrm{O}_2}$ is the total energy of an oxygen molecule (obtained
in a spin-polarized calculation).

Fig.~\ref{fig:fm} shows the oxygen vacancy formation energy in oxygen-reduced BaTiO$_{3-\delta}$ and oxygen-reduced LiNbO$_{3-\delta}$ as a
function of oxygen vacancy concentration. Red points and blue
triangles are the formation energies of oxygen vacancy in
BaTiO$_{3-\delta}$ and LiNbO$_{3-\delta}$, respectively.  The blue
cross is a reference value from previous
works~\cite{choi2011electronic}.

We use different supercells to test different concentrations of oxygen
vacancies. The formation energy of an oxygen vacancy does not have a
strong dependence on vacancy concentration, implying that the vacancy
concentration is low enough such that vacancy-vacancy interaction
is negligible.  The formation energy of a charge neutral oxygen
vacancy in LiNbO$_{3-\delta}$ is larger than that in
BaTiO$_{3-\delta}$ by about 0.9 eV per vacancy. This is reasonable considering
the fact that the gap of LiNbO$_3$ is larger than that of
BaTiO$_3$ by about 1 eV (see Fig.~\ref{fig:ferro-dos}).  While the
formation energy of oxygen vacancy in LiNbO$_{3-\delta}$ is higher,
oxygen vacancy has been widely observed in LiNbO$_{3-\delta}$ in
experiments~\cite{li2015defect,deleo1988electronic,smyth1983defects,sweeney1983oxygen}.

\section{Conclusion}

In conclusion, we perform first-principles calculations to study the
possible co-existence of conduction electrons and polar distortions in
$n$-doped BaTiO$_3$ and $n$-doped LiNbO$_3$, using both rigid-band
modeling and more realistic supercell calculations.  From rigid-band
modeling, we find that upon electron doping, cation displacements in
BaTiO$_3$ are quickly reduced and completely vanish at a critical
electron concentration of 0.1~$e$/f.u.. In contrast, in $n$-doped
LiNbO$_3$, Li-O and Nb-O displacements are significantly larger than
cation displacements in $n$-doped BaTiO$_3$, and more importantly they
are much more robust and can persist even at a concentration of
0.3~$e$/f.u..  In $n$-doped LiNbO$_3$, Li-O displacements decay more
slowly than Nb-O displacements, while in $n$-doped BaTiO$_3$, Ba-O and
Ti-O displacements decay approximately at the same rate.  From
supercell calculations (using oxygen vacancy as electron donors), we
find that in BaTiO$_{3-\delta}$ with $\delta = 4.2\%$, cation
displacements in BaTiO$_3$ are almost completely suppressed, which is
consistent with the result of rigid-band modeling. Conduction
electrons on Ti atoms are uniformly distributed, underlying rigid-band
calculations.  On the other hand, the results of oxygen-reduced
LiNbO$_{3-\delta}$ ($\delta = 4.2\%$) from supercell calculations are
more complicated than rigid-band calculations. Substantial polar
displacements $\delta_{\textrm{Li-O}}$ and $\delta_{\textrm{Nb-O}}$
occur throughout the supercell, but strong variations are found in
conduction electron distribution. This implies that the rigid-band approximation
should be used with caution in simulating electron doped LiNbO$_3$
and related oxides.

Our work indicates that electron doping of LiNbO$_3$-type
ferroelectrics is a simple and feasible approach to approximately
creating the rare polar metallic phase. Incorporating doped
ferroelectric semiconductors (in particular LiNbO$_3$-type
ferroelectrics) into devices may lead to new functionality and
applications.

\begin{acknowledgements}
  Hanghui Chen is supported by the National Natural Science Foundation of
  China under project number 11774236, Shanghai Pujiang Talents Program
  (Grant No. 17PJ1407300) and Seed grant of NYU-ECNU Research Institute
  of Physics. Yue Chen and Chengliang Xia are supported by
  Research Grants Council of Hong Kong under project numbers
  17200017 and 17300018, and the National Natural
  Science Foundation of China under project number 11874313.
  NYU Shanghai HPC and HKU-ITS provide computational resources.
\end{acknowledgements}

\bibliographystyle{apsrev}


\pagebreak

\begin{center}

\textbf{Supplementary Materials\\
Coexistence of polar displacements and conduction in doped
ferroelectrics: an \textit{ab initio} comparative study}
\end{center}

\setcounter{section}{0}
\setcounter{equation}{0}
\setcounter{figure}{0}
\setcounter{table}{0}
\setcounter{page}{1}
\renewcommand{\theequation}{S\arabic{equation}}
\renewcommand{\thefigure}{S\arabic{figure}}
\renewcommand{\bibnumfmt}[1]{[S#1]}
\renewcommand{\citenumfont}[1]{S#1}

\section{Details of crystal structures}

\begin{figure}[ht!]
\includegraphics[scale=0.45]{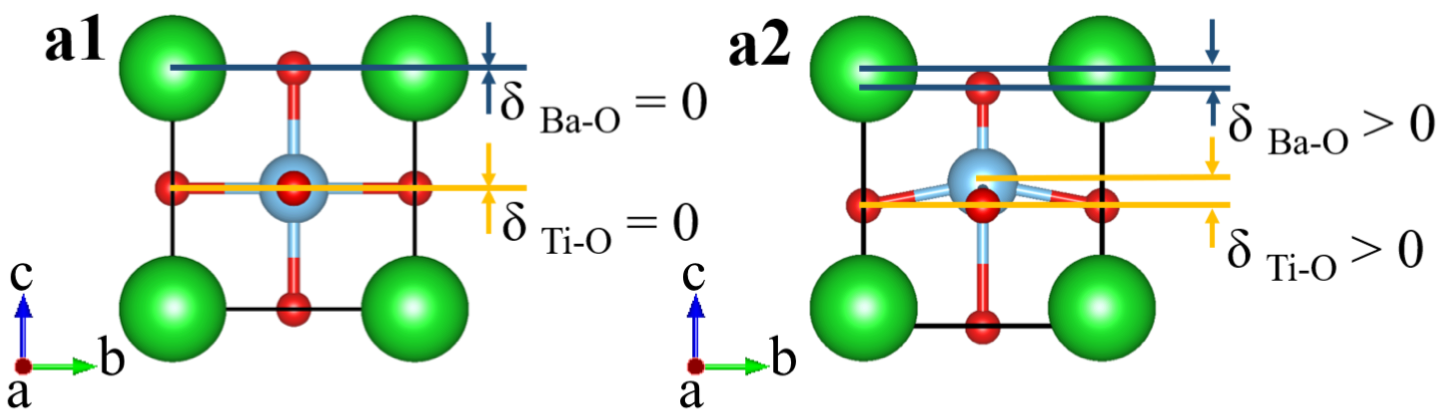}
\caption{Crystal structure of BaTiO$_3$ in rigid-band calculations. \textbf{a1}) $Pm\overline{3}m$. \textbf{a2}) $P4mm$.}
\label{fig:bto}
\end{figure}

Fig.~\ref{fig:bto} shows the crystal structure of BaTiO$_3$. Panel
\textbf{a1} shows zero Ba-O and Ti-O displacements in centrosymmetric
BaTiO$_3$ (space group $Pm\overline{3}m$), and it is the reference
state. Panel \textbf{a2} shows non-zero Ba-O and Ti-O displacements in
non-centrosymmetric BaTiO$_3$ (space group $P4mm$). Ti-O displacement
$\delta_{\textrm{Ti-O}}$ is defined as:
\begin{equation}
\delta_{\textrm{Ti-O}}= z_{\textrm{Ti}} - \frac{1}{4}\sum_{i=0}^4z_{\textrm{O}_i}
\end{equation}     
where $z_{\textrm{O}_i}$ is the $z$ position of the four nearest O atoms
around a given Ti atom in $xy$ plane. We note that in 119-atom
cell, Ti\#18 and Ti\#23 that are closest to
the oxygen vacancy have only three nearest O atoms.

\begin{figure}[t!]
\includegraphics[scale=0.55]{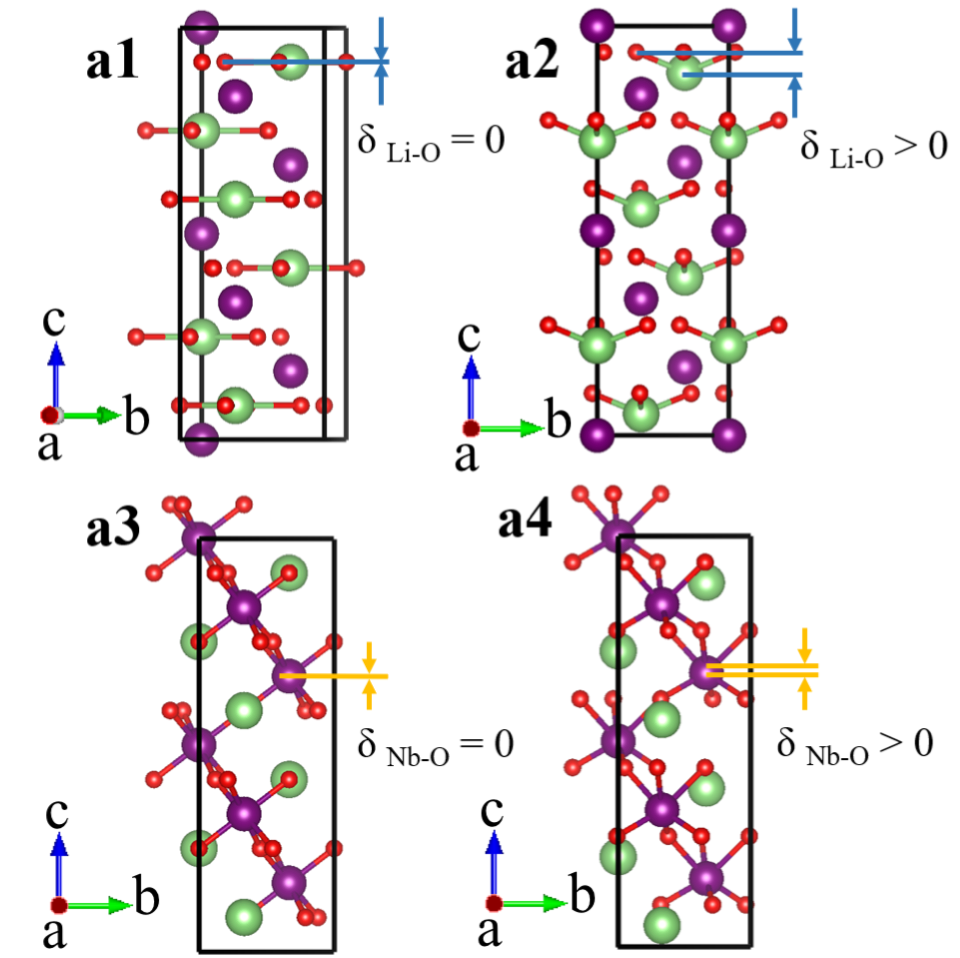}
\caption{Crystal structure of LiNbO$_3$ in rigid-band calculations.
  \textbf{a1}) and \textbf{a3}) $R\overline{3}c$.
  \textbf{a2}) and \textbf{a4}) $R3c$. 
\label{fig:lno}}
\end{figure}

Fig.~\ref{fig:lno} shows the crystal structure of LiNbO$_3$.
We show different bonding because we need different bondings to define $\delta_{\textrm{Li-O}}$
and $\delta_{\textrm{Nb-O}}$. To define $\delta_{\textrm{Li-O}}$, we need the
tetrahedron formed by one Li atom and three O atoms. For $\delta_{\textrm{Nb-O}}$, we need the octahedron with six oxygen atoms enclosing one Nb atom.
Panels \textbf{a1} and \textbf{a3} show zero Li-O and Nb-O displacements in
centrosymmetric LiNbO$_3$ (space group $R\overline{3}c$), and it is
the reference state. Panels \textbf{a2} and \textbf{a4} show non-zero
Li-O and Nb-O displacements in non-centrosymmetric LiNbO$_3$ (space
group $R3c$). Li-O displacement $\delta_{\textrm{Li-O}}$ is defined as:
\begin{equation}
\delta_{\textrm{Li-O}}= z_{\textrm{Li}} - \frac{1}{3}\sum_{i=0}^3 z_{\textrm{O}_i}
\end{equation}   
where $Z_{\textrm{O}_i}$ is $z$ position of the plane that is formed by the
three nearest O atoms for a given Li atom. We note that in a 119-atom cell,
Li\#4 that is closest to the oxygen vacancy has only two nearest O atoms.
Nb displacement $\delta_{\textrm{Nb-O}}$  is defined as:
\begin{equation}
  \delta_{\textrm{Nb-O}}= z_{\textrm{Nb}} - \frac{1}{6}\sum_{i=0}^6 z_{\textrm{O}_i}
\end{equation} 
where $z_{\textrm{O}_i}$ is $z$ position of the six nearest O atoms
around a given Nb atom. We note that in a 119-atom cell, Nb\#12 and
Nb\#24 that are closest to the oxygen vacancy have only five nearest O
atoms.

\newpage
\clearpage

\section{B\lowercase{a}-O cation displacements in oxygen reduced
  B\lowercase{a}T\lowercase{i}O$_{3-\delta}$}

\begin{figure}[!h]
\includegraphics[scale=0.3]{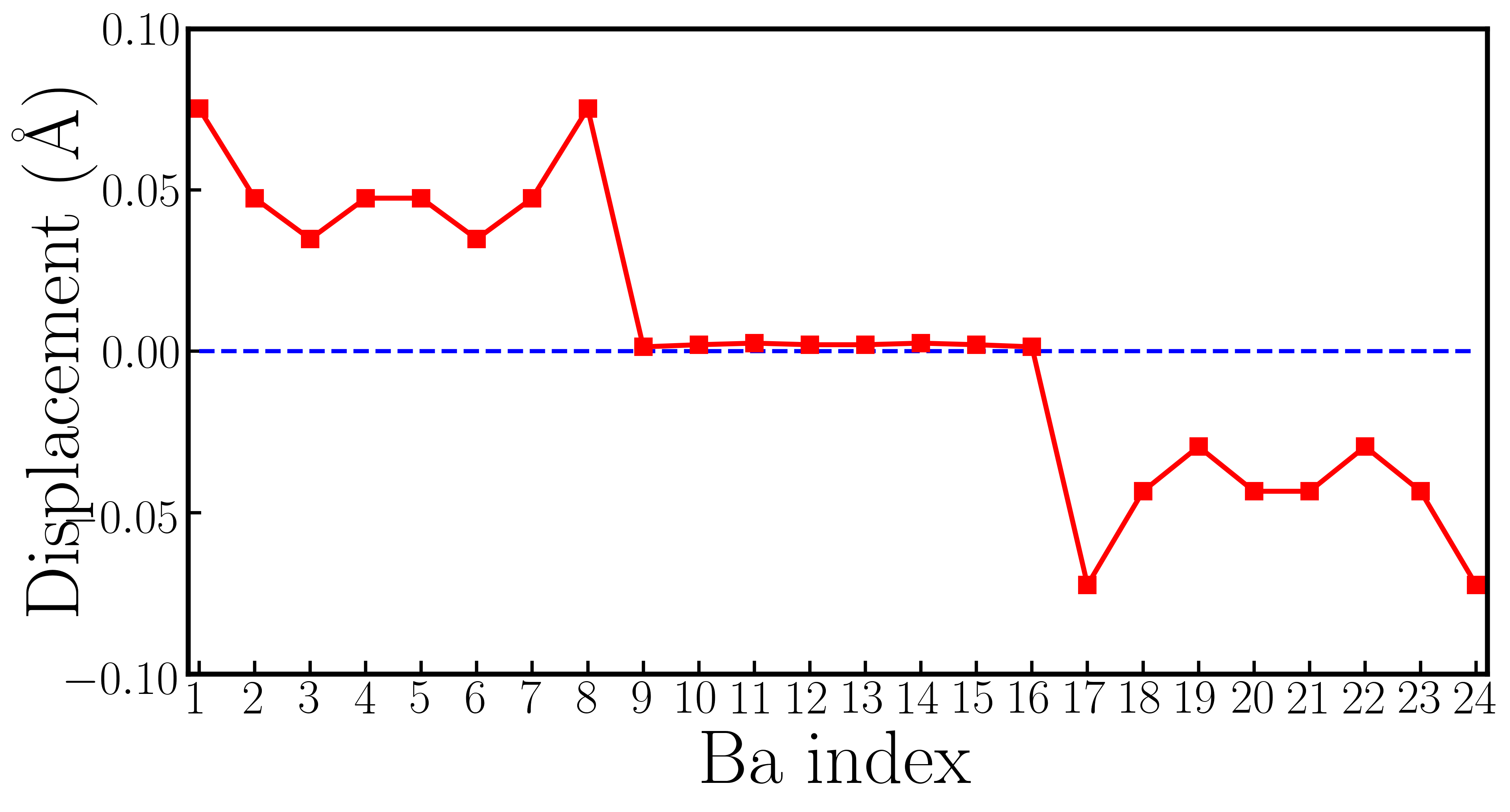}
\caption{Displacements along the $z$-axis of each Ba atom in oxygen reduced
  BaTiO$_{3-\delta}$ from our supercell calculations.} 
\label{fig:Ba-O}
\end{figure}

FIG.~\ref{fig:Ba-O} shows the Ba displacements along the $z$-axis in
the 119-atom BaTiO$_3$ supercell with an oxygen vacancy. It is noticed
that there is no net average Ba displacement in electron-doped
BaTiO$_3$ from our supercell calculations. The polar displacements on
Ba atoms \#1 and \#24 (\#8 and \#17 etc.) are of the same magnitude
but opposite in direction. These are caused by the presence of the
oxygen vacancy.

\newpage
\clearpage

\section{Comparison of LDA and PBE\lowercase{sol} calculations}

\begin{figure}[ht!]
\includegraphics[scale=0.4]{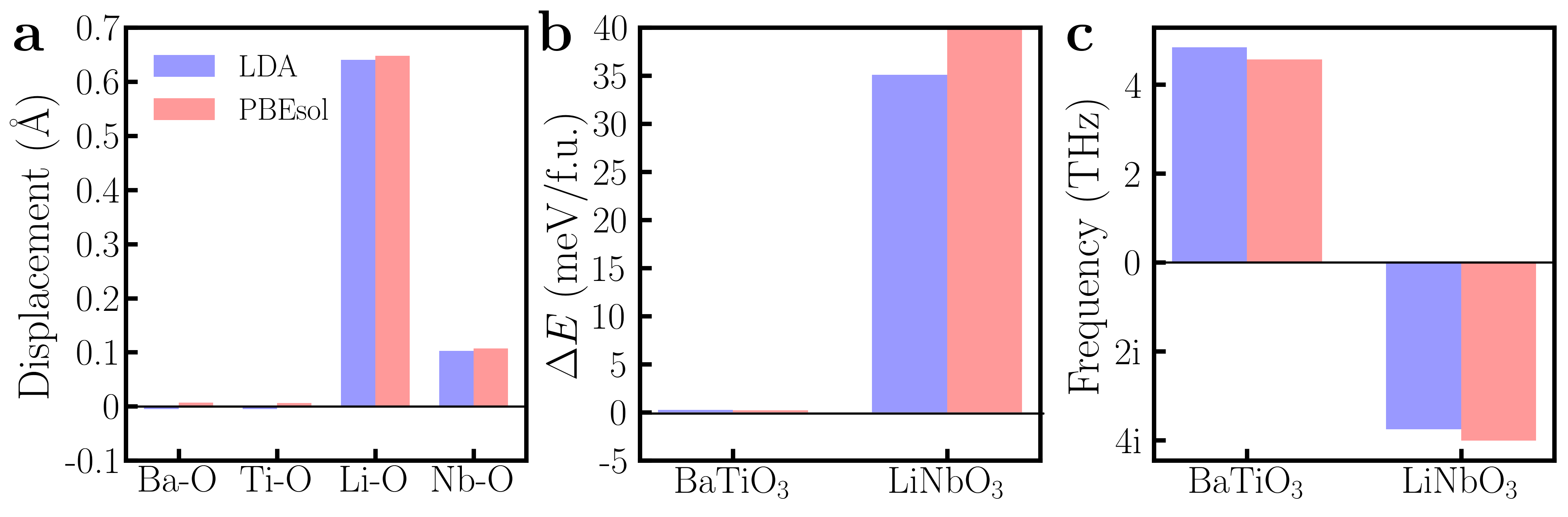}
\caption{Comparison of LDA and PBEsol calculations on the key results.
  Panel \textbf{a}) Ba-O, Ti-O, Li-O and Nb-O cation displacements in
  non-centrosymmetric structures of BaTiO$_3$ and LiNbO$_3$ with
  0.3$e$/f.u. doping concentration. Panel \textbf{b}) Energy difference between
  the centrosymmetric and non-centrosymmetric structures of BaTiO$_3$ and
  LiNbO$_3$ with 0.3$e$/f.u. doping concentration.
  Panel \textbf{c}) phonon
  frequency of the zone-center ferroelectric mode of cubic BaTiO$_3$ and
  $R\bar{3}c$ LiNbO$_3$ with 0.3$e$/f.u. doping concentration.}
\label{fig:PBEs-LDA}
\end{figure}

We check our key results using PBEsol and find no significant changes.
The comparison between LDA and PBEsol results is shown in
FIG.~\ref{fig:PBEs-LDA}.  Ba-O and Ti-O displacements are completely
suppressed upon 0.3$e$/f.u. doping, while Li-O and Nb-O displacements
are still significant at the same level of electron doping. At
0.3$e$/f.u. doping, the centrosymmetric and non-centrosymmetric
structures are essentially the same for BaTiO$_3$ but are distinct for
LiNbO$_3$. Our results are robust against different exchange
correlation.

\newpage
\clearpage

\section{Comparison of two type electrons in oxygen redueced
  B\lowercase{a}T\lowercase{i}O$_{3-\delta}$ and
  L\lowercase{i}N\lowercase{b}O$_{3-\delta}$}

\begin{figure}[ht!]
\includegraphics[scale=0.3]{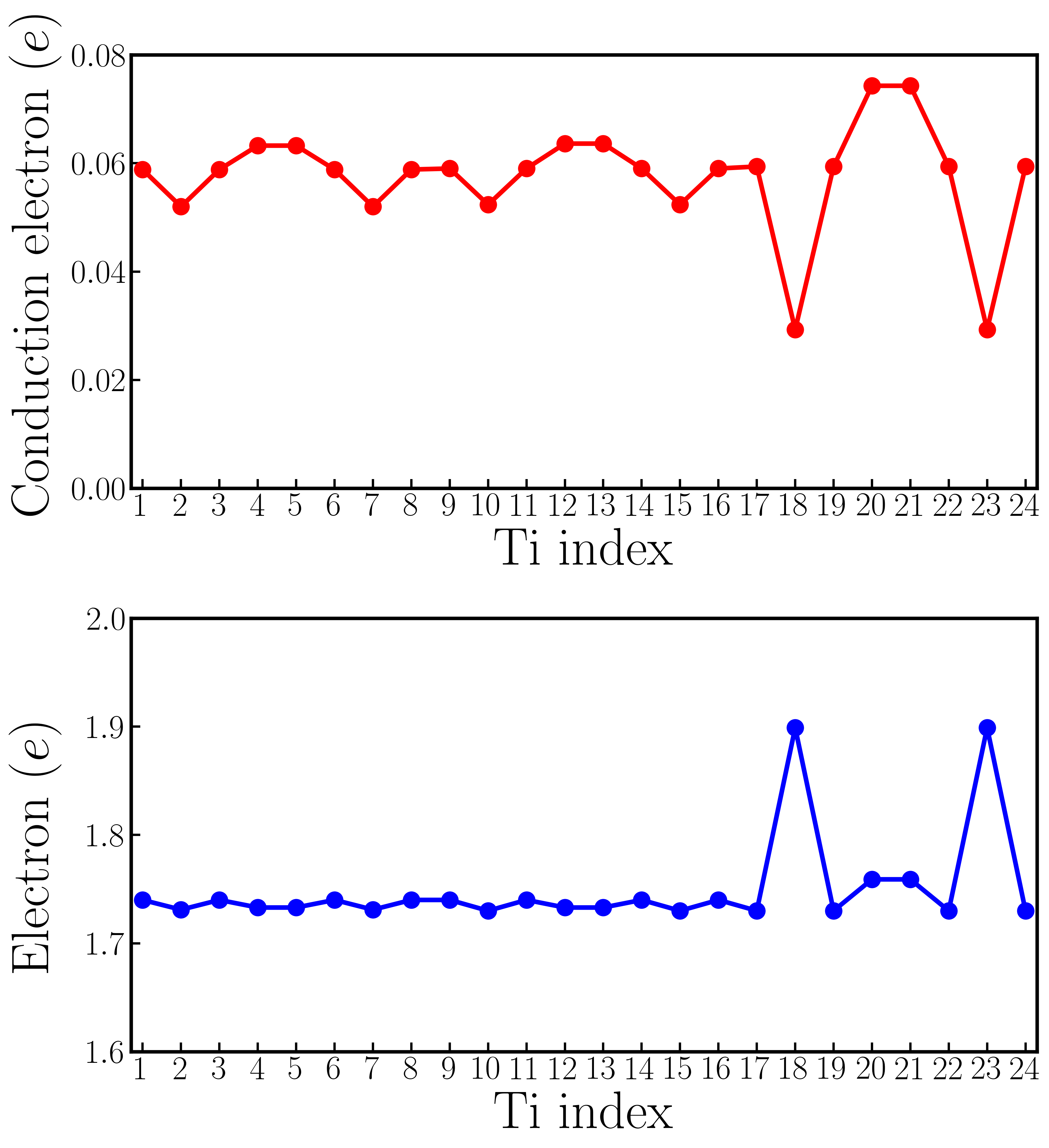}
\caption{\textbf{a}) The number of conduction electrons in Ti-$d$ states
  on each Ti atom by integrating Ti-$d$ states from the band gap to the
  Fermi level. \textbf{b}) The number of atomic Ti-$d$ electrons of each
  Ti atom by integrating Ti-$d$ electrons in a spherical region. Both results
  are obtained from a 119-atom BaTiO$_3$ supercell calculation.}
\label{fig:BaTiO3-e}
\end{figure}

Fig.~\ref{fig:BaTiO3-e} compares two types of Ti-$d$ electrons in
119-atom BaTiO$_3$ supercell with an oxygen vacancy.  The upper panel
shows the Ti-$d$ conduction electrons which are obtained by
integrating the Ti-$d$ states from the band gap to the Fermi
level. The lower panel shows the atomic Ti-$d$ electrons that are
obtained by integrating charge density in a spherical region. Atomic
Ti-$d$ electrons include both Ti-$d$ electrons in the conduction bands
(which are above the band gap) and Ti-$d$ electrons in the valence
bands (which are below the band gap). The Ti-$d$ electrons in the
valence bands are due to strong hybridization between Ti-$d$ and O-$p$
states~\cite{marianetti2004role}. From Fig.~\ref{fig:BaTiO3-e}, we
find that indeed the two nearest-neighbor Ti atoms have the largest
number of Ti-$d$ electrons. However, not all these electrons are in
the conduction bands.

\begin{figure}[t]
\includegraphics[scale=0.3]{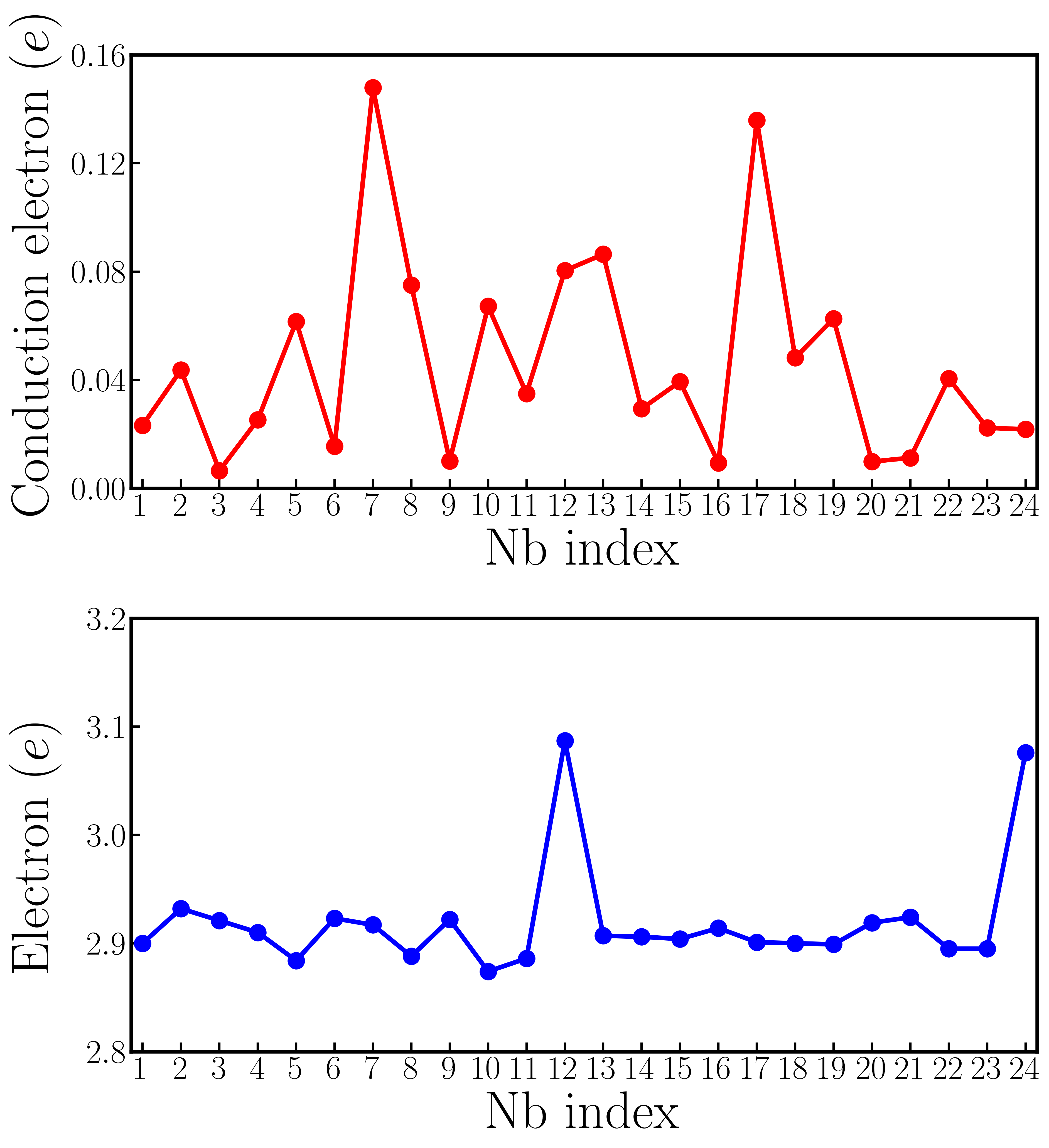}
\caption{\textbf{a}) The number of conduction electrons in Nb-$d$ states
  on each Nb atom by integrating Nb-$d$ states from the band gap to the
  Fermi level. \textbf{b}) The number of atomic Nb-$d$ electrons of each
  Nb atom by integrating Nb-$d$ electrons in a spherical region. Both results
  are obtained from a 119-atom LiNbO$_3$ supercell calculation.}
\label{fig:LiNbO3-e}
\end{figure}

Fig.~\ref{fig:LiNbO3-e} compares two-types of Nb-$d$ electrons in
119-atom LiNbO$_3$ supercell with an oxygen vacancy. Electron
distribution in LiNbO$_3$ is similar to that in BaTiO$_3$. Nb \#12 and
Nb\#24 have the most Nb-$d$ electrons (Fig.~\ref{fig:LiNbO3-e}
\textbf{b}), but their conduction electrons are not necessarily the
most (Fig.~\ref{fig:LiNbO3-e} \textbf{a}).

\newpage
\clearpage

\section{Calculations of oxygen reduced B\lowercase{a}T\lowercase{i}O$_{3-\delta}$ using LDA+$U$ or larger supercell}

\begin{figure}[ht!]
\includegraphics[scale=0.4]{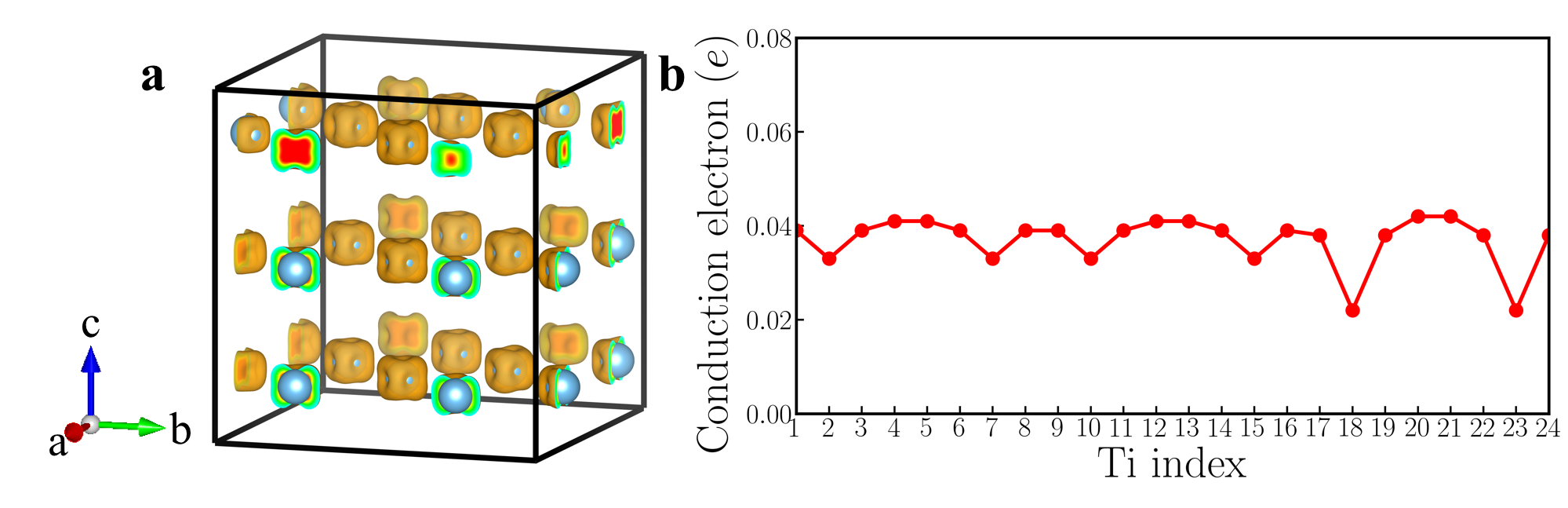}
\caption{\label{fig:LDAU}LDA+$U$ supercell calculations of electron
  doped BaTiO$_3$.  Panel \textbf{a)} an iso-value surface of
  conduction electron distribution in oxygen reduced
  BaTiO$_{3-\delta}$ with $\delta = 4.2\%$. Panel \textbf{b)}
  conduction electrons on each Ti atom in the 119-atom cell by
  integrating Ti-$d$ states from band gap to the Fermi level.}
\label{fig:Ti-d-LDAU}
\end{figure}

\begin{figure}[ht!]
\includegraphics[scale=0.3]{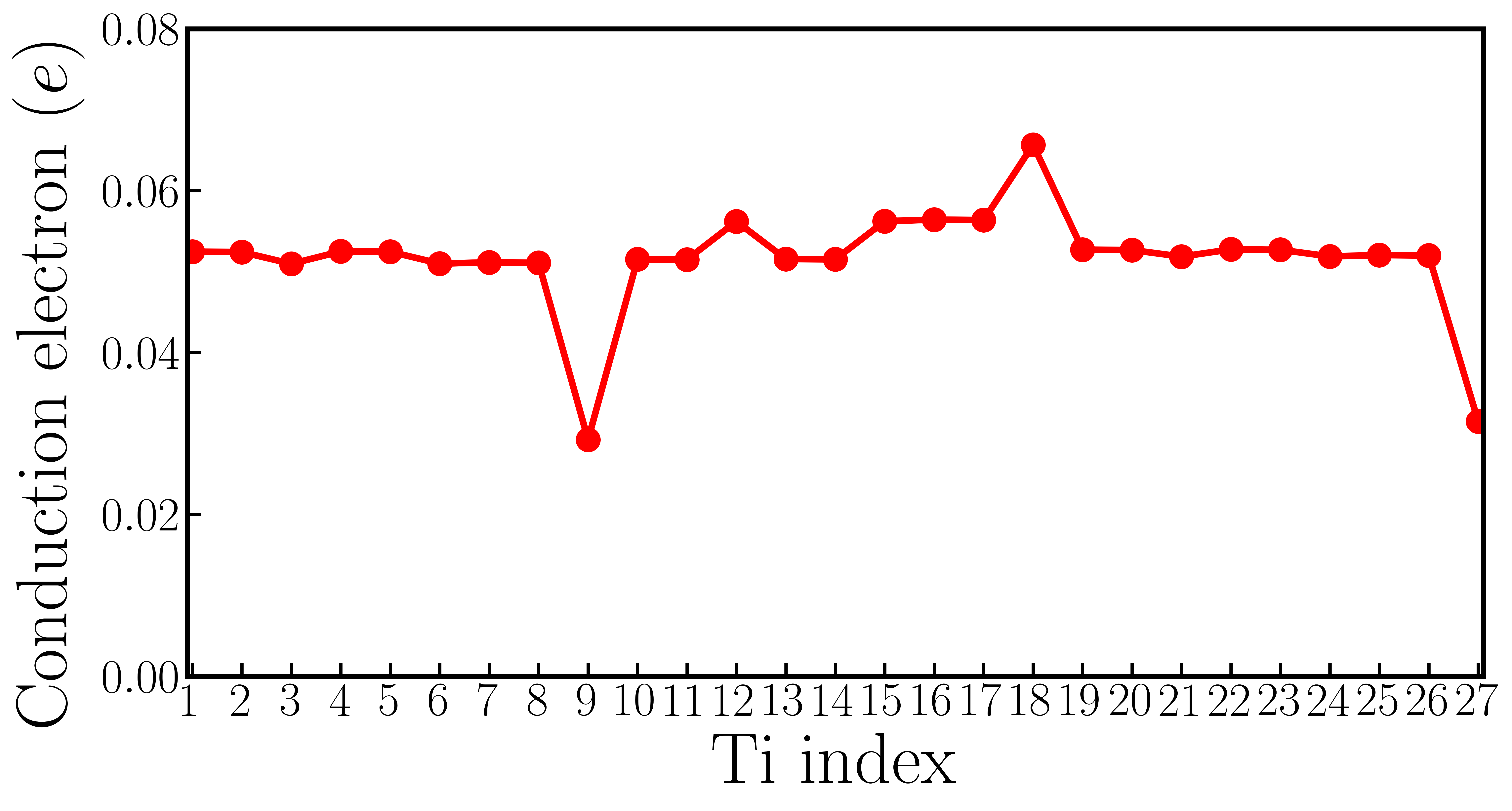}
\caption{\label{fig:BaTiO3-ti-134}Conduction electron on each Ti atom
  in the 134-atom BaTiO$_3$. The number of conduction electron on each
  Ti atom is obtained by integrating Ti-$d$ states from band gap to
  the Fermi level.  }
\end{figure}

We check oxygen reduced BaTiO$_3$ using LDA+$U$ method (with $U_{\textrm{Ti}}$
= 4 eV). The new results are shown in Fig.~\ref{fig:Ti-d-LDAU}. Panel
\textbf{a} shows the iso-value surface of Ti-$d$ conduction electrons
on Ti sites and panel \textbf{b} shows the number of Ti-$d$ conduction
electrons on each Ti atom in oxygen reduced BaTiO$_{3-\delta}$. From LDA+$U$
calculations, we find Ti-$d$ conduction electrons are still
homogeneously distributed.

We also test the cell size by calculating
conduction electrons on each Ti atom in the 134-atom
BaTiO$_3$ supercell. The new result is shown in
Fig.~\ref{fig:BaTiO3-ti-134}. Conduction electron distribution in Ti-$d$
states is homogeneous in a 134-atom BaTiO$_3$ cell, similar to 119-atom
cell calculation.

\newpage
\clearpage

\section{Calculations of oxygen redueced L\lowercase{i}N\lowercase{b}O$_{3-\delta}$ with a higher doping concentration}

\begin{figure}[ht!]
\includegraphics[scale=0.35]{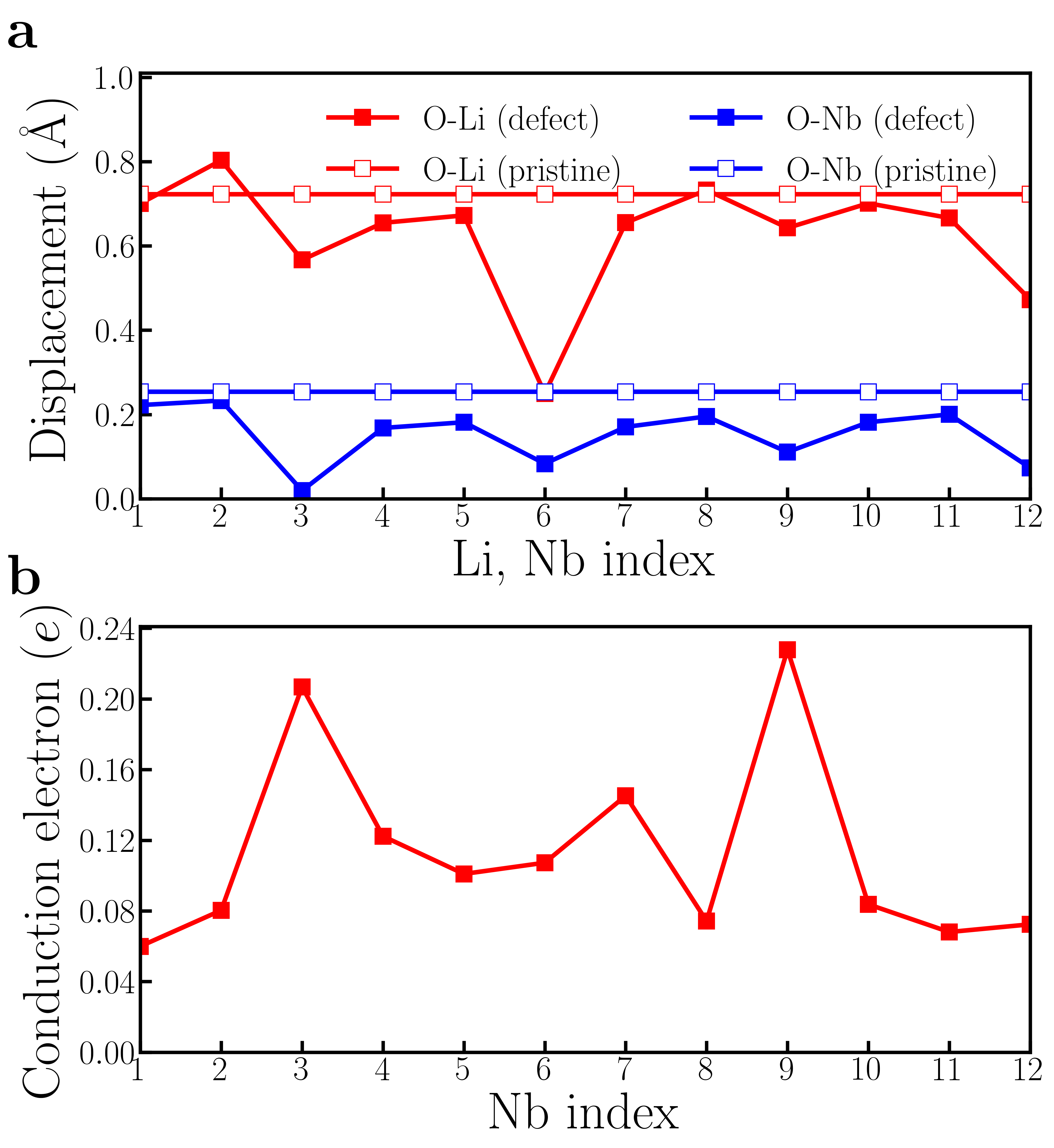}
\caption{Panel \textbf{a)} Polar displacements of each Li (red square)
  and Nb atom (blue square) in oxygen reduced LiNbO$_{3-\delta}$ with
  $\delta = 8.4\%/$f.u.. The solid squares are calculated from the
  59-atom cell. The open squares are bulk $\delta_{\textrm{Li-O}}$
  and $\delta_{\textrm{Nb-O}}$ of pristine LiNbO$_3$. Panel
  \textbf{b)} Conduction electrons on each Nb atom in the 59-atom
  cell. }
\label{fig:lno-59}
\end{figure}

FIG.~\ref{fig:lno-59}\textbf{a} shows $\delta_{\textrm{Li}-\textrm{O}}$
for each Li atom and $\delta_{\textrm{Nb}-\textrm{O}}$ for each Nb
atom in oxygen reduced LiNbO$_{3-\delta}$ with $\delta =
8.4\%/$f.u. FIG.~\ref{fig:lno-59}\textbf{b} shows the number of conduction
electrons on each Nb atom in 59-atom LiNbO$_3$ supercell. The number
of conduction electrons on each Nb atom is obtained by integrating
Nb-d states from band gap to the Fermi level. Li\#4, Nb\#6 and Nb\#12
are closest to the oxygen vacancy in 59-atom LiNbO$_3$ cell. The
minimum conduction electron in 59-atom LiNbO$_3$ is $0.06e$, which is
much larger than that in 119-atom LiNbO$_3$ supercell, only $0.0065e$.

\newpage
\clearpage

\section{Oxygen reduced LiNbO$_{3-\delta}$ with two oxygen vacancies}

\begin{figure}[ht!]
\includegraphics[scale=0.1]{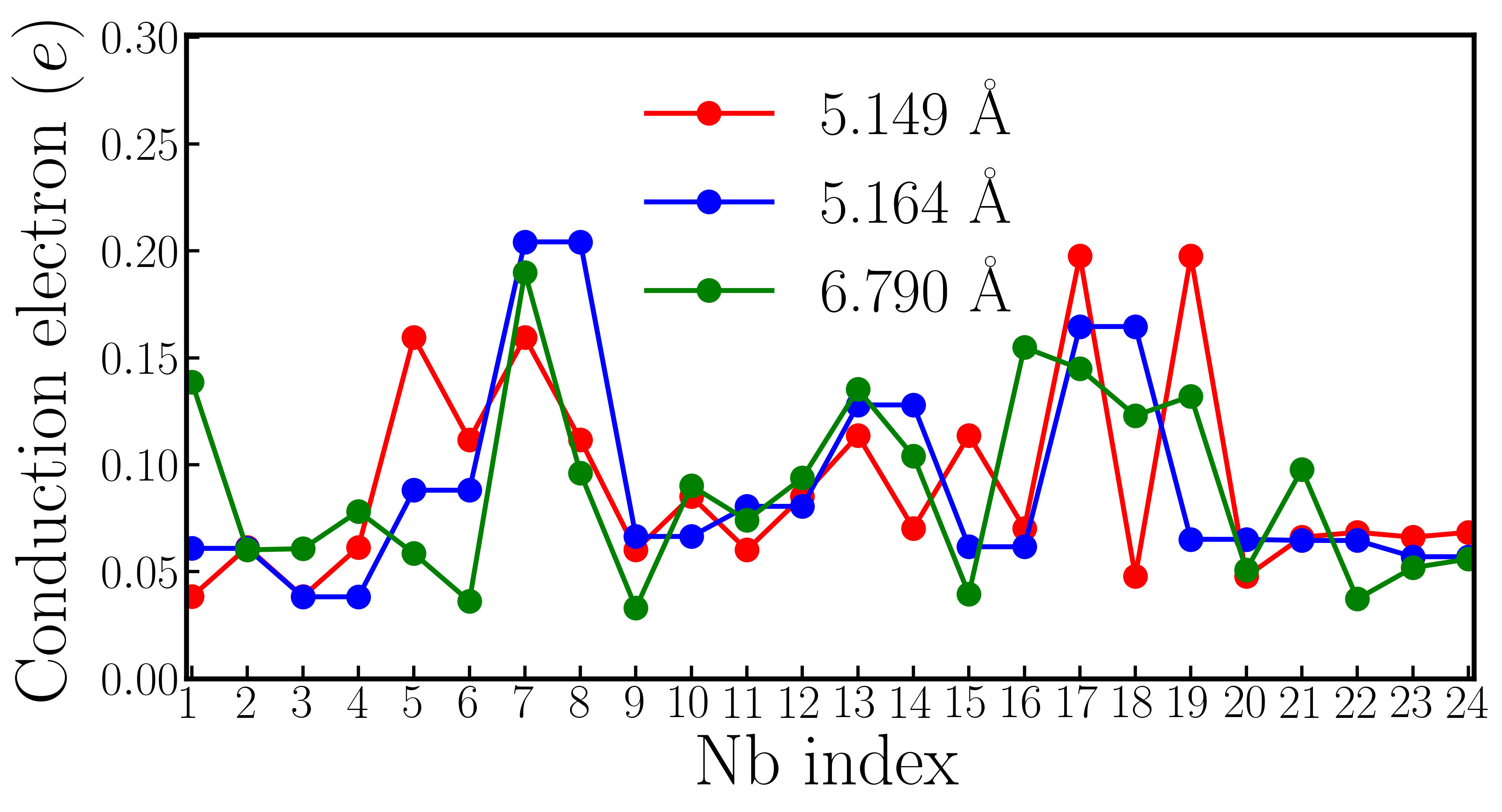}
\caption{Distribution of conduction electrons in Nb-$d$ states from
  three different configurations of a 118-atom cell that includes two
  oxygen vacancies.}
\label{fig:lno-118}
\end{figure}

We introduce two oxygen vacancies into a 120-atom LiNbO$_3$ supercell
and study conduction electron distribution in these cells with
different vacancy distances. These results are shown in
FIG.~\ref{fig:lno-118}. For all three separations, conduction electrons
on Nb atoms are inhomogeneously distributed.

\newpage
\clearpage


\end{document}